\documentclass[fleqn,usenatbib]{mnras}

\usepackage{newtxtext,newtxmath}

\usepackage[T1]{fontenc}

\DeclareRobustCommand{\VAN}[3]{#2}
\let\VANthebibliography\thebibliography
\def\thebibliography{\DeclareRobustCommand{\VAN}[3]{##3}\VANthebibliography}

\usepackage{graphicx}
\usepackage{amsmath}
\usepackage{multirow}

\newcommand{\bpf}{{\texttt{BpF\:}}}
\newcommand{\rsp}{{\texttt{RSP\:}}}
\newcommand{\hecto}{Hectoshelle@MMT}

\newcommand{\alfabar}{\bar{\alpha}}

\title[Empirical convective parameters for RRc models]{Temperature dependent convective parameters for RRc 1D-models}

\author[Kovács, Nuspl \& Szabó]{
Gábor B. Kovács,$^{1,2,3}$\thanks{E-mail: g.kovacs@astro.elte.hu}
János Nuspl,$^{2,3}$
Róbert Szabó$^{2,3,4}$
\\
$^{1}$ELTE Eötvös Loránd University,  Department of Astronomy  1117, P\'azm\'any P\'eter s\'et\'any 1/A, Budapest, Hungary \\
$^{2}$Konkoly Observatory, Research Centre for Astronomy and Earth Sciences, E\"otv\"os Lor\'and Research Network (ELKH), MTA Centre of Excellence\\ H-1121 Budapest, Konkoly Thege Mikl\'os \'ut 15-17, Hungary\\
$^{3}$MTA CSFK Lend\"ulet Near-Field Cosmology Research Group,  H-1121 Budapest, Konkoly Thege Mikl\'os \'ut 15-17, Hungary\\
$^{4}$ELTE Eötvös Loránd University, Institute of Physics, 1117, P\'azm\'any P\'eter s\'et\'any 1/A, Budapest, Hungary
}

\date{Accepted 2023 September 6. Received 2023 August 16; in original form 2023 June 8}

\pubyear{2023}

\begin{document}
\label{firstpage}
\pagerange{\pageref{firstpage}--\pageref{lastpage}}
\maketitle

\begin{abstract}
Nonlinear pulsation modeling of classical variable stars is among the first topics which were developed at the beginning of the computational era. Various developments were made, and many questions were answered in the past 60 years, and the models became more complex, describing  the genuinely 3D convection in a single dimension. Despite its successes, the recent public availability of the MESA Radial Stellar Pulsations (MESA RSP) module and the emerging results from multidimensional codes made clear that the 8 free convective parameters, unique to these models, together with the underlying physical models need calibration. This could be done by comparing them against  multi-dimensional codes, but before that, it is important to scrutinize the free parameters of the 1D codes using observations.
This is a follow-up work of our previous calibration on the convective parameters of the Budapest-Florida and MESA RSP pulsation codes for RRab stars. In this paper, we extend the previous calibration  to the RRc stars and the RR Lyrae stars in general.
We found that correlations of some of the parameters are present in RRc stars as well but have a different nature, while high-temperature RRc stars’ pulsation properties are very sensitive to the chosen parameter sets.
\end{abstract}


\defcitealias{KovacsGB2023}{Paper I}
\section{Introduction}

Nonlinear numerical modeling of classical pulsating stars (Cepheids, RR Lyrae stars) has a long history, which started with the first models of \cite{Christy1964} and culminated in the development of the various pulsation codes used today \citep{Bono1994,Yecko1998,lengyel}. In the meantime, the most important problems shifted from the driving mechanisms to the mode-selection problem \citep{Stellingwerf1982a,bpf-beat2002}. The physical formalism also became more complex, and the role of the convective processes in the outer layers of these stars became one of the main problems \citep{Deupree1977a}; for a quick walk-through of the development of the 1D nonlinear radial pulsation codes, we refer to \cite{KovacsGB2023}.

The theoretical description of the 1D convection theory also improved from the first attempts of time-dependent mixing length of \cite{Gough1977} and \cite{Unno1967} to the descriptions of \cite{Kuhfuss1986} and \cite{GW1992}. We suggest \cite{Baker1987} for a good review of the history of these theories.

The availability of running non-linear calculations was greatly enhanced by including the code of \cite{lengyel} into the MESA software package \citep{Paxton2019} as the \rsp\ module. Meanwhile, other groups used other codes to successfully model Cepheid stars \citep{Marconi2013b,Marconi2013a,Marconi2015,KellerWood2006}. The availability of these codes allows computing large grids of models, so for example, \cite{Susmita2021} have investigated convective and metallicity effects on BL Her model grids, while \cite{Kurbah2023} studied phase-dependent Cepheid period-luminosity relations using model grids, with different parameter sets.

The usage of different convective parameter sets is very common in these studies. This is because there is no overall calibration of the convective parameters currently. In the case of the \rsp code, there are only prescriptions \citep{Paxton2019}. It is well known that there should be no existing simple parameter set that suits every pulsator due to the limitations of the theory and the vast differences in the stellar structure of different types of stars in the instability strip \citep{lengyel}. Although \cite{bpf-beat2002} gave parameter sets for Cepheids and RR Lyrae stars for the Budapest-Florida code (\bpf\ hereafter) \citep{Yecko1998}, most studies in this field did not include stellar parameter dependence of the convective parameters,
 while there is evidence in the case of the pulsation code of \cite{Bono1994} that convective parameters may be affected by effective temperatures \citep{DiCriscienzo2004}.

Recently we have done a calibration for these parameters for the \rsp and \bpf codes in the case of the RRab stars  \citep[][hereafter Paper I]{KovacsGB2023}. In our \citetalias{KovacsGB2023}, we concluded that there is a degeneracy between the scale parameter of the eddy viscosity pressure $\alfabar_\nu$ and the dissipation efficiency parameter $\alfabar_d$, which is dependent on the effective temperature. We also showed that a discrepancy arises between the synthetic light and radial velocity (RV) curves around the blue edge of the instability strip. In the current work, we extend this  effort towards RRc stars in an attempt to find a general RR Lyrae parameter set (if such exists), which would help any large-scale modeling efforts requiring extensive model grid computations and will also provide a way into the transition applying multidimensional codes.

\section{Observational data}

 Albeit radial velocity and multi-band light-curve can be fitted with pulsation models simultaneously to derive stellar parameters \citep{DiFabrizio2002,Natale2008,Marconi2013a,Marconi2013b,Marconi2015}, this approach includes further assumptions regarding the atmosphere of the star which is not modeled by these codes \citepalias{KovacsGB2023}, while using only radial velocity curves we can study the dynamical structure more thoroughly. For this reason, we use only RV curves of \cite{Jurcsik2017} for calibration and V light curves from \cite{Jurcsik2015} only for comparison. 
We use these data
of selected first overtone RR Lyrae stars from the M3 globular cluster based on their RV curve coverage \citepalias{KovacsGB2023}. We use the cleaned and equidistantly re-sampled RV curves calculated from the \hecto\ \citep{Szentgyorgy2011}\ measurements with 1 km/s errors; we also show light curves for comparison, which were measured simultaneously with the \hecto\ measurements with the 60/90 Schmidt telescope at Konkoly Observatory \citep{Jurcsik2015}.

The value of the projection factor (the multiplier factor which connects the measured radial velocities to the actual pulsating velocities of the star and has a value between 1.0 and 1.5) is central in evaluating the radial velocity curves. There is an ongoing debate about whether this p-factor has a period dependence or not \citep[][and references therein]{Molinaro2012,Marconi2013b,Trahin2021}, which can be caused by structural differences among Cepheids, but among RR Lyrae stars the 
intrinsic stellar parameters span a much smaller parameter range.
In this regard, we choose the value of p-factor to $p=1.34 \pm 0.07$, the same as in \citetalias{KovacsGB2023}, and consider the uncertainty of the factor in the error propagation of the convective parameters. 

We derived the measured stellar parameters in the same way as in \citetalias{KovacsGB2023}. We have used bolometric corrections of \citet{Torres2010}, interstellar cluster reddening from \citet{Schlafly2011}, and the Baade-Wesselink distances from \citet{Jurcsik2017} to determine the bolometric luminosities of the stars.

\begin{table}
    \centering
    \caption{The used stars from \citet{Jurcsik2017}, and the selected input model parameters.}
    \label{tab:stars}
    \begin{tabular}{c|ccccccl}
        Star & 	Period$^\star$ & $T_{\textrm{eff}}^\star$ &	$L_{bol}^{\star\star}$ & $M^{\rm a}$ & $X^{\rm b}$ & $Z^{\rm b}$\\
          ID & 	[d] & [K] &	$[L_{\odot}]$ & [$M_\odot$] &  & \\        
        \hline
        v056&  	0.329598	&$7074$&	$44.7 \pm 2.8$ & 0.5636 & 0.76 & 0.0005\\
        v086& 	0.292656	&$7348$&	$46.2 \pm 2.5$ & 0.58154 & 0.71 & 0.0005\\
        v097& 0.334997&$7108$&    $47.7 \pm 2.8$ & 0.58339 & $0.75$ & $0.0006$\\
        v107& 	0.309026&$7259$&	$47.4 \pm 2.7$ & 0.58881 & 0.72 & 0.0006\\
        \multicolumn{7}{l}{\footnotesize $^\star$: directly adopted from \citet{Jurcsik2017}, error is $\pm 50\,\rm K$}\\
        \multicolumn{7}{l}{\footnotesize $^{\star\star}$: calculated from mean magnitudes of \citet{Jurcsik2015}}\\
        \multicolumn{7}{l}{ \footnotesize $^{\rm a}$: calculated from LNA model interpolation, error is $\pm 0.1\,M_\odot$}\\
        \multicolumn{7}{l}{ \footnotesize $^{\rm b}$: derived from the non-linear fits, $X,Z$ errors are $\pm 0.03$ and $\pm 0.0001$}\\
        \multicolumn{7}{l}{\ \ \ \ respectively}
    \end{tabular}
\end{table}

\section{Models and the fitting method}

We aim at determining the eight convective parameters of the non-linear radial pulsation models \citep{GW1992,Kuhfuss1986} from radial velocity observations in the case of RRc stars in a similar way as it has been recently done for RRab stars in \citetalias{KovacsGB2023}.

We use the Budapest-Florida code \citep{Yecko1998} and the MESA Radial Stellar Pulsation module \citep{Paxton2019} for the  convective parameter fitting. These two codes are very similar to each other; their main difference is the handling of negative buoyancy effects \citepalias{KovacsGB2023}. For a one-to-one comparison, see our \citetalias{KovacsGB2023} and also the original papers of \citet{Yecko1998} and \citet{lengyel}. 

These types of models have 5 input parameters intrinsic to the star: the stellar mass ($M$), the bolometric luminosity ($L$), the effective temperature ($T_{\rm eff}$), the hydrogen  ($X$) and a metal mass fraction ($Z$). Some of these 
parameters are directly measurable ($L$,$T_{\rm eff}$), while others can be determined indirectly by the models. As in \citetalias{KovacsGB2023}, we determine the pulsating mass by interpolating the period from linear models and $X$ and $Z$ by finding the best-fit values from non-linear calculations.

Our fitting procedure remains unchanged from \cite{KovacsGB2023}. Briefly, it means that we  run model grids for the dissipation efficiency $\alfabar_d$ and eddy viscosity parameter  $\alfabar_\nu$, as these two parameters have the largest effect on the observed RV curves, and they are also degenerate \citep{KovacsGB2023}. The turbulent source ($\alfabar_s$) and convective flux ($\alfabar_c$) parameters have stronger effects on the synthetic RV and LC curves of first overtones, hence instead of independent fitting \citepalias[as in][]{KovacsGB2023} we fit these parameters together.

We are fitting the full radial velocity (RV) curve (in contrast with \citetalias{KovacsGB2023}, where minimum phase was omitted), which weakens the overall fit but helps to avoid nonphysical artificially strong dissipation fronts   in the models, that would cause very strong secondary light curve features e.g. flare-like spikes.

\section{Results}

\begin{table}
    \centering
    \caption{Parameters of the $\alfabar_\nu$-$\alfabar_d$ regression: $\alfabar_d=a+ b\alfabar_\nu$}
    \label{tab:regressions}
    \begin{tabular}{c|ccc|ccc}
         & \multicolumn{3}{|c|}{\bpf} & \multicolumn{3}{|c|}{\rsp} \\
        Star & $a$ & $b$ & $R^2$ & $a$ & $b$ & $R^2$ \\
        \hline
        v056 & -16.35 & 381.65 & 0.9661 & 0.16886 & 61.3950 & 0.9645 \\
        v086 & -24.59 & 802.76 & 0.8682 & 0.22024 & 130.582 & 0.9590 \\
        v097 & -23.59 & 374.39 & 0.9475 & 0.85470 & 58.4948 & 0.9698 \\
        v107 & -14.01 & 532.11 & 0.8649 & 1.43412 & 87.7540 & 0.9896 \\

    \end{tabular}

\end{table}

\begin{figure}
    \centering
    \includegraphics[width=\columnwidth]{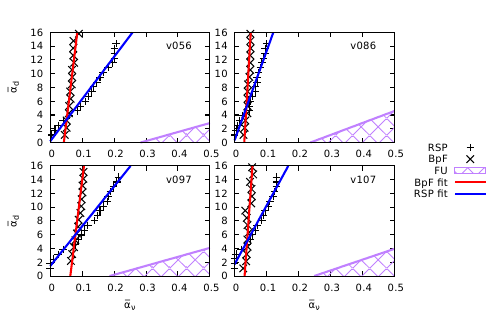}
    \caption{Correlation between the eddy viscosity ($\alfabar_\nu$, x axis) and turbulent dissipation ($\alfabar_d$, y axis) parameters. Crosses are the best fits of the \bpf\ code, while plus signs are the best fits of the \rsp\ code, and the red and blue lines are the regression lines for the \bpf\ and \rsp\, respectively. The purple crossed area refers to parameters where the pulsation switches to fundamental mode and is damped.}
    \label{fig:regressions}
\end{figure}

\begin{figure}
    \centering
    \includegraphics[width=0.95\columnwidth]{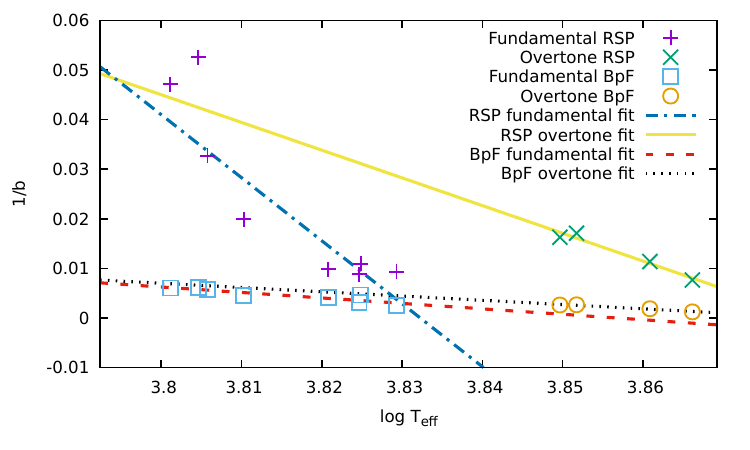}
    \caption{ Correlation between the effective temperature and the slope of the $\alpha_\nu(\alpha_d)$ functions for both pulsation modes and codes as well. Every point is the $1/b$ value from Table \ref{tab:regressions} corresponding to a single star and one of the codes (see legend), while the x axis is the logarithm of the effective temperature. The fundamental mode values are from \citetalias{KovacsGB2023}.  The lines are the correlations corresponding to the pulsation codes and modes.
    One can see that the RRab and RRc correlations differ for both codes, but it is more prominent in the case of the \rsp\ code.
    }
    \label{fig:slopes}
\end{figure}

\begin{figure}
    \centering
    \includegraphics[width=0.95\columnwidth]{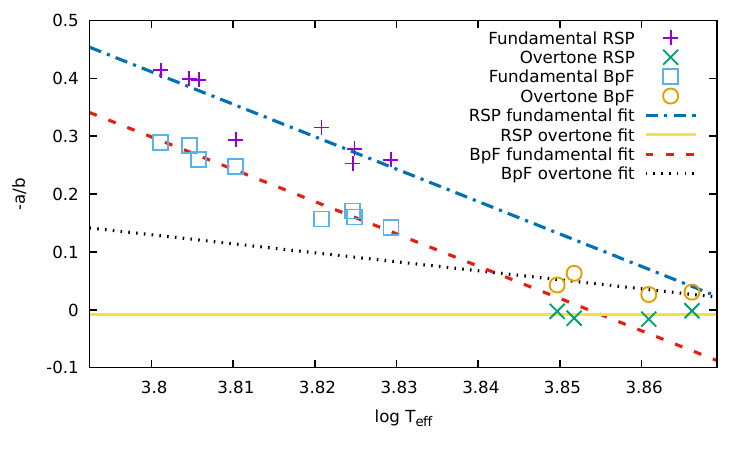}
    \caption{Same as Fig.~\ref{fig:slopes}. but for the interception ($-a/b$ from Table \ref{tab:regressions}.) of the $\alpha_\nu(\alpha_d)$ functions.}
    \label{fig:interceptions}
\end{figure}

\begin{figure}
    \centering
    \includegraphics[width=0.95\columnwidth]{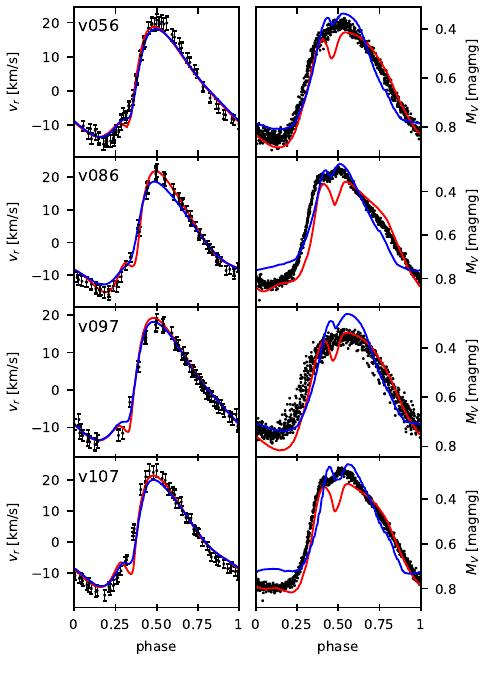}
    \caption{Best fitting results for the studied M3 RRc stars. Left panel: radial velocities, right panel: absolute V magnitude. \bpf results marked with the red line, \rsp\ is the blue line. Black dots mark observations.}
    \label{fig:best-fits}
\end{figure}

The RRc stars show a similar degeneracy between the dissipation ($\alfabar_d$) and eddy viscosity ($\alfabar_\nu$) parameters as  was observed in the case of RRab stars \citep{KovacsGB2023} (see Fig.~\ref{fig:regressions}), but the regression parameters have a different correlation with the effective temperature which is shown in Figs.~\ref{fig:slopes}. and \ref{fig:interceptions}. We show our best-fit RV curves alongside the light curves as a reference in Fig.~\ref{fig:best-fits}. We see good agreement in the amplitudes, albeit secondary light curve features are too large in some cases, especially in the \bpf\ code, which can be attributed to the sensitivity of the features to the underlying convection models \citep{Marconi2017rev}.

 In addition to the known effects \citepalias{KovacsGB2023} of $\alfabar_\nu$ and $\alfabar_d$, we found that the convective flux ($\alfabar_c$) and source parameters ($\alfabar_s$) have stronger effects on the first overtone stars, and we found that on a $\alfabar_c-\alfabar_s$ grid, the parameters show a degeneracy of hyperbolic shape, for each star. We can see this also by recalculating this grid for the RRab stars from \citetalias{KovacsGB2023}. This hyperbole has an interception with the $\alfabar_s=\alfabar_c$ line, so we can further reduce the number of free parameters by one.

We present the details of these results for both codes separately below.

\begin{table}
    \centering
    \caption{Convective flux parameters for the two codes choosing $\alfabar_c=\alfabar_s$}
    \begin{tabular}{ccc}
    Star & \rsp & \bpf \\
    \hline
    v086 & $0.2572 \pm 0.02$ & $0.1418 \pm 0.03$ \\
    v097 & $0.4361 \pm 0.02$ & $0.1880 \pm 0.05$ \\
    v056 & $0.3591 \pm 0.04$ & $0.2093 \pm 0.03$ \\
    v107 & $0.3716 \pm 0.08$ & $0.1658 \pm 0.03$ \\
    \end{tabular}
    
    \label{tab:c_s}
\end{table}

\subsection{\bpf\ results}

In the case of the \bpf\ the $\alfabar_d-\alfabar_\nu-T_{\rm eff}$ correlation is the following:

\begin{multline}
    \alfabar_\nu = (-1.5568 \pm 1) \log T_{\rm eff} -( 0.0859 \pm 0.01 ) \alfabar \log T_{\rm eff}\\
    +( 0.3336 \pm 0.04) \alfabar_d + ( 6.0454 \pm 3.92) 
\end{multline}
     The rms of the fit is $0.010$.

The $\alfabar_c=\alfabar_s$ parameters (\ref{tab:c_s}) show a correlation with temperature ($R^2=0.9568$), while RRab parameters don't show this feature. But due to the large errors and the low number of points, one can use a generally constant value of $0.17\pm 0.02$ for both parameters, which is within the errors of the previous values of RRab stars. $\alfabar_t$ and $\alfabar_p$ have little effect on actual RV and light curves.

\subsection{\rsp\ results}

In the case of the \rsp\ code, the correlation of the $\alfabar_\nu-\alfabar_d-T_{\rm eff}$ is as follows:

\begin{multline}
    \alfabar_\nu = (-0.5591 \pm 0.078) \alfabar_d \log T_{\rm eff}\\
    +( 2.16595 \pm 0.03) \alfabar_d - ( 0.0088 \pm 0.0038) 
\end{multline}
 with an rms of $0.027$.

The $\alfabar_c=\alfabar_s$ parameters are weakly correlated ($R^2\approx0.5$) and differ from the RRab case. The changes introduced on the RV curves are small while having stronger effects on the light curves. Because of this,  we also considered the light curve shape in the fitting process, and this way, the mean $\alfabar_c=0.35 \pm 0.06$, which is greater than the RRab value of $\alfabar_c=0.30 \pm 0.07$ but they match within the errorbars. Because of this, we can suggest a general value of $\alfabar_c=\alfabar_s=0.32 \pm 0.07$. The $\alfabar_t$ (turbulent flux) parameter has a small effect, and its best value is around $0.61$.

\subsection{General RR Lyrae parameter set}

\begin{figure}
    \centering
    \includegraphics[width=\columnwidth]{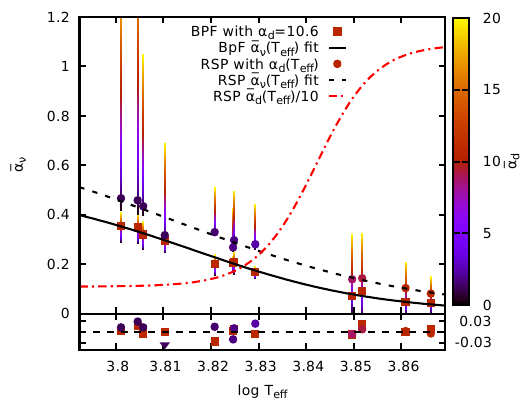}
    \caption{General $\alfabar_\nu(T_{\rm eff})$ functions fitted to the RR Lyrae stars. Colored lines show the individual $\alfabar_\nu(\alfabar_d)$ functions in the $\log T_{\rm eff}-\alfabar_\nu$ plane ($\alfabar_d\in[0,20]$), the color denotes the $\alfabar_d$ value. Filled squares: \bpf\ values for $\alfabar_d=10.8$, black solid curve: fitted sigmoid of the \bpf. Filled circles: results of the \rsp\ code. Here $\alfabar_d$ also changes to reach the best-fit value, which is shown by the red dashed-dotted curve (for better visibility, we show $\alfabar_d$/10). The black dashed curve is the fitted sigmoid to the \rsp\ values. The lower panel show residuals where the color codes are as the $\alfabar_d$ value of the point and the triangle shows outlier points.}
    \label{fig:Overall_fit}
\end{figure}

We also derived a general set for the eddy viscosity ($\alfabar_\nu$) and dissipation ($\alfabar_d$) parameters, which can be seen in Fig.~\ref{fig:Overall_fit}. For this purpose, we have chosen the logistic function in the form of:
\begin{equation}
    \label{eq:logistic}
    f(T_{\rm eff})=\frac{a}{\exp [{b(T_{\rm eff}-c)}]+1}+d
\end{equation}

This function can quickly converge to $d$ and $a+d$, which is ideal for avoiding unphysical parameters (negative and too high $\alfabar$-s).

In the \bpf\ case, we could get a good fit with $\alfabar_d=10.8$, while in the case of the \rsp\ code, it was necessary for $\alfabar_d$ to change with a logistic function, too. We present these parameters in Table~\ref{tab:logistic}.

\begin{table}
    \centering
    \begin{center}
    
    \caption{Parameters of function Eq.~\ref{eq:logistic}, in different cases.}
    \centerline{\begin{tabular}{c|c|c|c|c|c}
         Case & $a$ & $b$ & $c$ & $d$ & rms \\
        \hline
        \bpf\ $\alfabar_\nu$ & $0.526\pm0.008$ & $0.0033\pm0.0002$ & $6525\pm10$ & $0.005$ &0.015 \\
      \rsp\ $\alfabar_\nu$ & $0.775\pm0.02$ & $0.0024\pm0.0002$ & $6444\pm10$ & $0.01$ &0.02 \\
      \rsp\ $\alfabar_d$ & $9.7979$ & $-0.01$ & $6950$ &  1.089 & N/A\\
    \end{tabular}}
    \label{tab:logistic}
    \end{center}
\end{table}

\section{Discussion}

\begin{figure}
    \centering
    \includegraphics[width=\columnwidth]{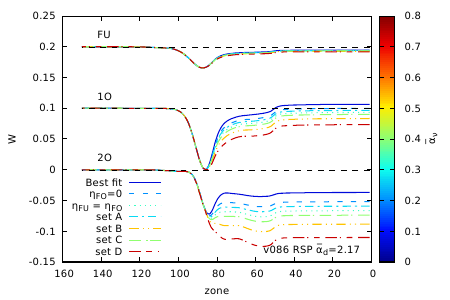}
    \caption{Cumulative work integrals for different $\alfabar_\nu$ parameters in the case of the v086 RRc star calculated by the \rsp\ code, shifted to fit on a single figure. Color code shows the value of the $\alfabar_\nu$ parameter. From top to bottom: fundamental mode, 1st overtone, 2nd overtone. Black dashed line corresponds to zero. }
    \label{fig:workintegrand}
\end{figure}

\begin{figure}
    \centering
    \includegraphics[width=\columnwidth]{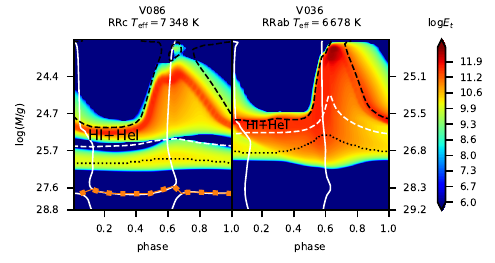}
    \caption{Changing the turbulent energy profile throughout the pulsations in an RRc (left panel, $T_{\rm eff}=7348$ K) and in an RRab model (right panel, $T_{\rm eff}=6678$ K.) star calculated by the \rsp\ code. X axis is the pulsation phase, Y axis is the logarithm of the mass coordinate from the surface. Color denotes the strength of the turbulent energy. White lines are  the zero velocity isocurves. HI and HeI partial ionization zones overlaps, the boundaries of this overlapped zone are denoted by black (outer boundary) and white (inner boundary) dashed curves. The HeII zone is narrow, we show it by the black dotted curve. The position of the nodal line in the RRc star is denoted by the orange dotted curve. Note that due to the phase differences the nodal line is moving and also briefly disappearing in certain phases, which is shown by the fact that the zero velocity curves are not contacting each other. RRc stars have different convective patterns: the HeII zone's convective region is separated from the HI region. In the case of the RRab stars, the two regions are merged. Towards lower temperatures, $e_t$ is also higher.}
    \label{fig:et}
\end{figure}

The fact that we arrived at different parameter sets for different pulsational modes can be understood since the convective processes influence mode selection, and in turn, the selected mode affects the stellar structure and transport processes. In practice, this means that the $\alfabar$ parameters depend not only on the physical parameters of the matter but also on the flow structure. We can see an example of this interaction in Fig.~\ref{fig:et}. Here we can see that in the case of the RRc stars, the flow itself helps to separate the two convective regions, while in the RRab case, the convective zones are connected independently from the temperature. The exact interaction between pulsation and convection is still a research subject, and it can be done using multi-dimensional pulsation codes \citep{kupka,Mundprecht2015,SPHERLS4}. 

As the mode selection problem can lead to disjoint sets of structural properties between pulsators \citep{bpf-drrlyr2004}, this different response to the parameters is understandable but also may question the validity of the mode-selection surveys. In spite of this, the \bpf\ code has more restricted parameter space, which strengthens these studies, while in the case of the \rsp\ code, set B and D are in the purple region of Fig.\ref{fig:regressions}, which means this parameter sets can lead to misleading results. Moreover, higher temperature RRc stars can be rendered pulsationally stable (non-pulsating) with parameter sets A and B as well. In Figure~\ref{fig:workintegrand} we can see the linear cumulative work integrals\footnote{The cumulative sum of the amount of work done by the sum of the pressure terms during one pulsational cycle calculated over the zones} of the star V086 ($T_{\rm eff}=7348$ K). We can see that the effect of the convective parameters increases for higher overtones. Altogether, we can assume that some features like the fundamental blue edge \citep{bpf-drrlyr2004} or the period-doubling phenomena \citep{Smolec2016} are probably more robust, while other features like RRd regions \citep{bpf-drrlyr2004,lengyel2} are more sensitive to the convective parameters.

The joint RR Lyrae parameter set has been chosen and calibrated to describe stars in the given temperature range $T\in [6000;7500]$ K, and due to the sigmoid functions, it has believable constraints outside of this range. Further work on calibration to Cepheid stars may reveal more details regarding the connection between the parameter space and intrinsic stellar parameters. The light and RV curves are insensitive to the turbulent flux and pressure ($\alfabar_t$,$\alfabar_p$) parameters, making some features smoother. These parameters have larger roles in mode selection, so they can be better calibrated by e.g., fundamental  blue edge or other features. 

\section{Conclusions}

We continued our previous work \citep{KovacsGB2023} on scrutinizing the convective parameter space of the 1D non-linear radial stellar pulsation codes, the Budapest-Florida and MESA Radial Stellar Pulsations. We used high-precision radial velocity measurements of four first overtone RR Lyrae stars from globular cluster M3 measured by \citet{Jurcsik2017}.

We fitted the convective parameters to these RV curves and found degeneracies between the parameters similar but also different to the RRab case. This different nature of the parameters can be interpreted as the effect of the different flow structures of the modes.
Nevertheless, the two (RRab and RRc) parameter sets can be approximated by sigmoid functions to reach a general RR Lyrae parameter set, which can be used for survey-like studies and also for comparison with multi-dimensional models. Some parameters have little effect on the observables, but mode-selection features can calibrate them. Our final results on these temperature-dependent parameter sets can be found in Table~\ref{tab:overall}.

We also found that the standard four-parameter set of the \rsp\ code \citep{Paxton2019} is inadequate to describe some of the RRc stars as they damp pulsations, while in some cases, sets B and D interfere with the mode selection process. We emphasize the need for more studies in this direction and warn against using parameter sets without sanity checks.

Numerical simulations in 2 and 3 dimensions are the next logical step in pulsation modeling. 
Our results give further basis to 1D-3D model comparisons \citep[e.g.][]{Mundprecht2015} by providing observationally calibrated parameters.

\begin{table}
    \centering
    \caption{Recommended parameter sets for RRab \citepalias{KovacsGB2023}, RRc stars, and generally RR Lyrae stars (this paper).}
    \begin{tabular}{cc|c|c|c|}
         &$\alfabar$ & RRab set & RRc set &  RR Lyr set \\
          \hline
        \parbox[t]{2mm}{\multirow{9}{*}{\rotatebox[origin=c]{90}{\rsp}}}
        &$\Bar{\alpha}_\Lambda$ & 1.5 &1.5 &1.5\\
        &$\bar{\alpha}_\nu$ &  $-8.65\log T_\textrm{eff} +33.38$ &$-6.085\log T_{\rm eff} +23.564$ & Table \ref{tab:logistic}.\\
        &$\bar{\alpha}_t$ &  $0.24 \pm 0.03$ & $0.61 \pm 0.05$ & $0.61\pm 0.05$\\
        &$\bar{\alpha}_p$ &  $2/3$ &$2/3$&$2/3$\\
        &$\bar{\alpha}_d$ &  $8/3\sqrt{2/3}$ &$10.883$& Table \ref{tab:logistic}.\\
        &$\bar{\alpha}_s$ & $0.31 \pm 0.07$ &$0.35\pm0.06$ & $0.32\pm 0.06$ \\
        &$\bar{\alpha}_c$ &  $0.27\pm 0.03$& $0.35\pm0.06$ & $0.32\pm 0.06$\\
        &$\bar{\alpha}_r$ & $0$ & $0$& $0$\\
        &$T_{\rm eff}$ & $< 7227$ K & $< 7455$ K & N/A\\
        \hline
        \hline
        \parbox[t]{2mm}{\multirow{9}{*}{\rotatebox[origin=c]{90}{\bpf}}}
        &$\Bar{\alpha}_\Lambda$  & 1.5 & 1.5 & 1.5\\
        &$\bar{\alpha}_\nu$ &  $-6.77 \log T_\textrm{eff} + 26.09$ & $-2.467\log T_{\rm eff} +9.581 $ & Table \ref{tab:logistic}.
\\
        &$\bar{\alpha}_t$ &  $0.2733$ & $0.2733$ & $0.2733$\\
        &$\bar{\alpha}_p$ &  $2/3$ & $2/3$ & $2/3$\\
        &$\bar{\alpha}_d$ &  $10.6$ & $10.6$ & $10.6$\\
        &$\bar{\alpha}_s$ &  $0.22 \pm 0.08$ & $0.17 \pm 0.02$ & $0.17$\\
        &$\bar{\alpha}_c$ &  $0.17\pm 0.02$ & $0.17\pm0.02$ & $0.17$\\
        &$\bar{\alpha}_r$ &  $0$ & $0$ & $0$\\
        &$T_{\rm eff} $& $<7141$ K& $<7650$ K & N/A
    \end{tabular}
    
    \label{tab:overall}
\end{table}

\section*{Acknowledgements}

This project has been supported by the Lend\"ulet Program  of the Hungarian Academy of Sciences, project No. LP2018-7/2022, the `SeismoLab' KKP-137523 \'Elvonal, OTKA projects K-129249 and NN-129075, as well as the MW-Gaia COST Action (CA18104).

On behalf of Project 'Hydrodynamical modeling of classical pulsating variables with SPHERLS' we are grateful for the usage of ELKH Cloud \citep[see][\url{https://science-cloud.hu/}]{H_der_2022} which helped us achieve the results published in this paper. 

\section*{Data Availability}

The observations used in this work are publicly available from the online material of \citet{Jurcsik2017}, and the \rsp\ numerical code is also available publicly as part of the \texttt{MESA} software \citep{Paxton2019}.

\bibliographystyle{mnras}
\bibliography{master}

\begin{thebibliography}{}
\makeatletter
\relax
\def\mn@urlcharsother{\let\do\@makeother \do\$\do\&\do\#\do\^\do\_\do\%\do\~}
\def\mn@doi{\begingroup\mn@urlcharsother \@ifnextchar [ {\mn@doi@}
  {\mn@doi@[]}}
\def\mn@doi@[#1]#2{\def\@tempa{#1}\ifx\@tempa\@empty \href
  {http://dx.doi.org/#2} {doi:#2}\else \href {http://dx.doi.org/#2} {#1}\fi
  \endgroup}
\def\mn@eprint#1#2{\mn@eprint@#1:#2::\@nil}
\def\mn@eprint@arXiv#1{\href {http://arxiv.org/abs/#1} {{\tt arXiv:#1}}}
\def\mn@eprint@dblp#1{\href {http://dblp.uni-trier.de/rec/bibtex/#1.xml}
  {dblp:#1}}
\def\mn@eprint@#1:#2:#3:#4\@nil{\def\@tempa {#1}\def\@tempb {#2}\def\@tempc
  {#3}\ifx \@tempc \@empty \let \@tempc \@tempb \let \@tempb \@tempa \fi \ifx
  \@tempb \@empty \def\@tempb {arXiv}\fi \@ifundefined
  {mn@eprint@\@tempb}{\@tempb:\@tempc}{\expandafter \expandafter \csname
  mn@eprint@\@tempb\endcsname \expandafter{\@tempc}}}

\bibitem[\protect\citeauthoryear{{Baker}}{{Baker}}{1987}]{Baker1987}
{Baker} N.~H.,  1987, in {Hillebrandt} W.,  {Meyer-Hofmeister} E.,  {Thomas}
  H.~C.,   {Kippenhahn} R.,  eds, Physical Processes in Comets, Stars and
  Active Galaxies. pp 105--124

\bibitem[\protect\citeauthoryear{{Bono} \& {Stellingwerf}}{{Bono} \&
  {Stellingwerf}}{1994}]{Bono1994}
{Bono} G.,  {Stellingwerf} R.~F.,  1994, \mn@doi [\apjs] {10.1086/192054},
  \href {https://ui.adsabs.harvard.edu/abs/1994ApJS...93..233B} {93, 233}

\bibitem[\protect\citeauthoryear{{Christy}}{{Christy}}{1964}]{Christy1964}
{Christy} R.~F.,  1964, \mn@doi [Reviews of Modern Physics]
  {10.1103/RevModPhys.36.555}, \href
  {https://ui.adsabs.harvard.edu/abs/1964RvMP...36..555C} {36, 555}

\bibitem[\protect\citeauthoryear{{Das}, {Kanbur}, {Smolec}, {Bhardwaj}, {Singh}
   \& {Rejkuba}}{{Das} et~al.}{2021}]{Susmita2021}
{Das} S.,  {Kanbur} S.~M.,  {Smolec} R.,  {Bhardwaj} A.,  {Singh} H.~P.,
  {Rejkuba} M.,  2021, \mn@doi [\mnras] {10.1093/mnras/staa3694}, \href
  {https://ui.adsabs.harvard.edu/abs/2021MNRAS.501..875D} {501, 875}

\bibitem[\protect\citeauthoryear{{Deupree}}{{Deupree}}{1977}]{Deupree1977a}
{Deupree} R.~G.,  1977, \mn@doi [\apj] {10.1086/154958}, \href
  {https://ui.adsabs.harvard.edu/abs/1977ApJ...211..509D} {211, 509}

\bibitem[\protect\citeauthoryear{{Di Criscienzo}, {Marconi}  \& {Caputo}}{{Di
  Criscienzo} et~al.}{2004}]{DiCriscienzo2004}
{Di Criscienzo} M.,  {Marconi} M.,   {Caputo} F.,  2004, \memsai, \href
  {https://ui.adsabs.harvard.edu/abs/2004MmSAI..75..190D} {75, 190}

\bibitem[\protect\citeauthoryear{{Di Fabrizio} et~al.,}{{Di Fabrizio}
  et~al.}{2002}]{DiFabrizio2002}
{Di Fabrizio} L.,  et~al., 2002, \mn@doi [\mnras]
  {10.1046/j.1365-8711.2002.05824.x}, \href
  {https://ui.adsabs.harvard.edu/abs/2002MNRAS.336..841D} {336, 841}

\bibitem[\protect\citeauthoryear{{Gehmeyr} \& {Winkler}}{{Gehmeyr} \&
  {Winkler}}{1992}]{GW1992}
{Gehmeyr} M.,  {Winkler} K. H.~A.,  1992, \aap, \href
  {https://ui.adsabs.harvard.edu/abs/1992A&A...253...92G} {253, 92}

\bibitem[\protect\citeauthoryear{{Geroux} \& {Deupree}}{{Geroux} \&
  {Deupree}}{2015}]{SPHERLS4}
{Geroux} C.~M.,  {Deupree} R.~G.,  2015, \mn@doi [ApJ]
  {10.1088/0004-637X/800/1/35}, \href
  {https://ui.adsabs.harvard.edu/abs/2015ApJ...800...35} {800, 35}

\bibitem[\protect\citeauthoryear{{Gough}}{{Gough}}{1977}]{Gough1977}
{Gough} D.~O.,  1977, \mn@doi [\apj] {10.1086/155244}, \href
  {https://ui.adsabs.harvard.edu/abs/1977ApJ...214..196G} {214, 196}

\bibitem[\protect\citeauthoryear{H{\'{e}}der et~al.,}{H{\'{e}}der
  et~al.}{2022}]{H_der_2022}
H{\'{e}}der M.,  et~al., 2022, \mn@doi [Inform{\'{a}}ci{\'{o}}s
  T{\'{a}}rsadalom] {10.22503/inftars.xxii.2022.2.8}, 22, 128

\bibitem[\protect\citeauthoryear{{Jurcsik} et~al.,}{{Jurcsik}
  et~al.}{2015}]{Jurcsik2015}
{Jurcsik} J.,  et~al., 2015, \mn@doi [\apjs] {10.1088/0067-0049/219/2/25},
  \href {https://ui.adsabs.harvard.edu/abs/2015ApJS..219...25J} {219, 25}

\bibitem[\protect\citeauthoryear{{Jurcsik} et~al.,}{{Jurcsik}
  et~al.}{2017}]{Jurcsik2017}
{Jurcsik} J.,  et~al., 2017, \mn@doi [\mnras] {10.1093/mnras/stx382}, \href
  {https://ui.adsabs.harvard.edu/abs/2017MNRAS.468.1317J} {468, 1317}

\bibitem[\protect\citeauthoryear{{Keller} \& {Wood}}{{Keller} \&
  {Wood}}{2006}]{KellerWood2006}
{Keller} S.~C.,  {Wood} P.~R.,  2006, \mn@doi [\apj] {10.1086/501115}, \href
  {https://ui.adsabs.harvard.edu/abs/2006ApJ...642..834K} {642, 834}

\bibitem[\protect\citeauthoryear{{Koll{\'a}th}, {Buchler}, {Szab{\'o}}  \&
  {Csubry}}{{Koll{\'a}th} et~al.}{2002}]{bpf-beat2002}
{Koll{\'a}th} Z.,  {Buchler} J.~R.,  {Szab{\'o}} R.,   {Csubry} Z.,  2002,
  \mn@doi [\aap] {10.1051/0004-6361:20020182}, \href
  {https://ui.adsabs.harvard.edu/abs/2002A&A...385..932K} {385, 932}

\bibitem[\protect\citeauthoryear{{Kov{\'a}cs}, {Nuspl}  \&
  {Szab{\'o}}}{{Kov{\'a}cs} et~al.}{2023}]{KovacsGB2023}
{Kov{\'a}cs} G.~B.,  {Nuspl} J.,   {Szab{\'o}} R.,  2023, \mn@doi [\mnras]
  {10.1093/mnras/stad884}, \href
  {https://ui.adsabs.harvard.edu/abs/2023MNRAS.521.4878K} {521, 4878}

\bibitem[\protect\citeauthoryear{{Kuhfuss}}{{Kuhfuss}}{1986}]{Kuhfuss1986}
{Kuhfuss} R.,  1986, \aap, \href
  {https://ui.adsabs.harvard.edu/abs/1986A&A...160..116K} {160, 116}

\bibitem[\protect\citeauthoryear{{Kupka} \& {Muthsam}}{{Kupka} \&
  {Muthsam}}{2017}]{kupka}
{Kupka} F.,  {Muthsam} H.~J.,  2017, \mn@doi [Living Reviews in Computational
  Astrophysics] {10.1007/s41115-017-0001-9}, \href
  {https://ui.adsabs.harvard.edu/abs/2017LRCA....3....1K} {3, 1}

\bibitem[\protect\citeauthoryear{{Kurbah}, {Deb}, {Kanbur}, {Das}, {Deka},
  {Bhardwaj}, {Randall}  \& {Kalici}}{{Kurbah} et~al.}{2023}]{Kurbah2023}
{Kurbah} K.,  {Deb} S.,  {Kanbur} S.~M.,  {Das} S.,  {Deka} M.,  {Bhardwaj} A.,
   {Randall} H.~R.,   {Kalici} S.,  2023, \mn@doi [\mnras]
  {10.1093/mnras/stad806}, \href
  {https://ui.adsabs.harvard.edu/abs/2023MNRAS.521.6034K} {521, 6034}

\bibitem[\protect\citeauthoryear{{Marconi}}{{Marconi}}{2017}]{Marconi2017rev}
{Marconi} M.,  2017, in European Physical Journal Web of Conferences. p. 06001,
  \mn@doi{10.1051/epjconf/201715206001}

\bibitem[\protect\citeauthoryear{{Marconi}, {Molinaro}, {Ripepi}, {Musella}  \&
  {Brocato}}{{Marconi} et~al.}{2013a}]{Marconi2013b}
{Marconi} M.,  {Molinaro} R.,  {Ripepi} V.,  {Musella} I.,   {Brocato} E.,
  2013a, \mn@doi [\mnras] {10.1093/mnras/sts197}, \href
  {https://ui.adsabs.harvard.edu/abs/2013MNRAS.428.2185M} {428, 2185}

\bibitem[\protect\citeauthoryear{{Marconi} et~al.,}{{Marconi}
  et~al.}{2013b}]{Marconi2013a}
{Marconi} M.,  et~al., 2013b, \mn@doi [\apjl] {10.1088/2041-8205/768/1/L6},
  \href {https://ui.adsabs.harvard.edu/abs/2013ApJ...768L...6M} {768, L6}

\bibitem[\protect\citeauthoryear{{Marconi} et~al.,}{{Marconi}
  et~al.}{2015}]{Marconi2015}
{Marconi} M.,  et~al., 2015, \mn@doi [\apj] {10.1088/0004-637X/808/1/50}, \href
  {https://ui.adsabs.harvard.edu/abs/2015ApJ...808...50M} {808, 50}

\bibitem[\protect\citeauthoryear{{Molinaro} et~al.,}{{Molinaro}
  et~al.}{2012}]{Molinaro2012}
{Molinaro} R.,  et~al., 2012, \mn@doi [\apj] {10.1088/0004-637X/748/1/69},
  \href {https://ui.adsabs.harvard.edu/abs/2012ApJ...748...69M} {748, 69}

\bibitem[\protect\citeauthoryear{{Mundprecht}, {Muthsam}  \&
  {Kupka}}{{Mundprecht} et~al.}{2015}]{Mundprecht2015}
{Mundprecht} E.,  {Muthsam} H.~J.,   {Kupka} F.,  2015, \mn@doi [\mnras]
  {10.1093/mnras/stv434}, \href
  {https://ui.adsabs.harvard.edu/abs/2015MNRAS.449.2539M} {449, 2539}

\bibitem[\protect\citeauthoryear{{Natale}, {Marconi}  \& {Bono}}{{Natale}
  et~al.}{2008}]{Natale2008}
{Natale} G.,  {Marconi} M.,   {Bono} G.,  2008, \mn@doi [\apjl]
  {10.1086/526518}, \href
  {https://ui.adsabs.harvard.edu/abs/2008ApJ...674L..93N} {674, L93}

\bibitem[\protect\citeauthoryear{Paxton et~al.,}{Paxton
  et~al.}{2019}]{Paxton2019}
Paxton B.,  et~al., 2019, \mn@doi [The Astrophysical Journal Supplement Series]
  {10.3847/1538-4365/ab2241}, 243, 10

\bibitem[\protect\citeauthoryear{{Schlafly} \& {Finkbeiner}}{{Schlafly} \&
  {Finkbeiner}}{2011}]{Schlafly2011}
{Schlafly} E.~F.,  {Finkbeiner} D.~P.,  2011, \mn@doi [\apj]
  {10.1088/0004-637X/737/2/103}, \href
  {https://ui.adsabs.harvard.edu/abs/2011ApJ...737..103S} {737, 103}

\bibitem[\protect\citeauthoryear{{Smolec}}{{Smolec}}{2016}]{Smolec2016}
{Smolec} R.,  2016, \mn@doi [\mnras] {10.1093/mnras/stv2868}, \href
  {https://ui.adsabs.harvard.edu/abs/2016MNRAS.456.3475S} {456, 3475}

\bibitem[\protect\citeauthoryear{{Smolec} \& {Moskalik}}{{Smolec} \&
  {Moskalik}}{2008a}]{lengyel}
{Smolec} R.,  {Moskalik} P.,  2008a, ActAA, \href
  {https://ui.adsabs.harvard.edu/abs/2008AcA....58..193S} {58, 193}

\bibitem[\protect\citeauthoryear{{Smolec} \& {Moskalik}}{{Smolec} \&
  {Moskalik}}{2008b}]{lengyel2}
{Smolec} R.,  {Moskalik} P.,  2008b, \actaa, \href
  {https://ui.adsabs.harvard.edu/abs/2008AcA....58..233S} {58, 233}

\bibitem[\protect\citeauthoryear{{Stellingwerf}}{{Stellingwerf}}{1982}]{Stellingwerf1982a}
{Stellingwerf} R.~F.,  1982, \mn@doi [\apj] {10.1086/160425}, \href
  {https://ui.adsabs.harvard.edu/abs/1982ApJ...262..330S} {262, 330}

\bibitem[\protect\citeauthoryear{{Szab{\'o}}, {Koll{\'a}th}  \&
  {Buchler}}{{Szab{\'o}} et~al.}{2004}]{bpf-drrlyr2004}
{Szab{\'o}} R.,  {Koll{\'a}th} Z.,   {Buchler} J.~R.,  2004, \mn@doi [\aap]
  {10.1051/0004-6361:20035698}, \href
  {https://ui.adsabs.harvard.edu/abs/2004A&A...425..627S} {425, 627}

\bibitem[\protect\citeauthoryear{{Szentgyorgyi} et~al.,}{{Szentgyorgyi}
  et~al.}{2011}]{Szentgyorgy2011}
{Szentgyorgyi} A.,  et~al., 2011, \mn@doi [\pasp] {10.1086/662209}, \href
  {https://ui.adsabs.harvard.edu/abs/2011PASP..123.1188S} {123, 1188}

\bibitem[\protect\citeauthoryear{Torres}{Torres}{2010}]{Torres2010}
Torres G.,  2010, \mn@doi [The Astronomical Journal]
  {10.1088/0004-6256/140/5/1158}, 140, 1158

\bibitem[\protect\citeauthoryear{{Trahin}, {Breuval}, {Kervella}, {M{\'e}rand},
  {Nardetto}, {Gallenne}, {Hocd{\'e}}  \& {Gieren}}{{Trahin}
  et~al.}{2021}]{Trahin2021}
{Trahin} B.,  {Breuval} L.,  {Kervella} P.,  {M{\'e}rand} A.,  {Nardetto} N.,
  {Gallenne} A.,  {Hocd{\'e}} V.,   {Gieren} W.,  2021, \mn@doi [\aap]
  {10.1051/0004-6361/202141680}, \href
  {https://ui.adsabs.harvard.edu/abs/2021A&A...656A.102T} {656, A102}

\bibitem[\protect\citeauthoryear{{Unno}}{{Unno}}{1967}]{Unno1967}
{Unno} W.,  1967, \pasj, \href
  {https://ui.adsabs.harvard.edu/abs/1967PASJ...19..140U} {19, 140}

\bibitem[\protect\citeauthoryear{{Yecko}, {Kollath}  \& {Buchler}}{{Yecko}
  et~al.}{1998}]{Yecko1998}
{Yecko} P.~A.,  {Kollath} Z.,   {Buchler} J.~R.,  1998, AAP, \href
  {https://ui.adsabs.harvard.edu/abs/1998A&A...336..553Y} {336, 553}

\makeatother
\end{thebibliography}

\bsp
\label{lastpage}
\end{document}